\documentclass[sigconf,natbib]{acmart}

\AtBeginDocument{%
  \providecommand\BibTeX{{%
    \normalfont B\kern-0.5em{\scshape i\kern-0.25em b}\kern-0.8em\TeX}}}
\usepackage[labelformat=simple]{subcaption}

\usepackage{enumitem}
\setlist{nosep}
\usepackage{bm}
\usepackage[linesnumbered,ruled]{algorithm2e}
\let\oldnl\nl
\newcommand{\nonl}{\renewcommand{\nl}{\let\nl\oldnl}}

\usepackage{algpseudocode}
\usepackage{balance}
\usepackage{titlesec}

\titleformat{\subsubsection}[runin]  
  {\bfseries\normalsize}            
  {\thesubsubsection}               
  {0.5em}                           
  {}                                

\usepackage{booktabs}
\usepackage{float}
\usepackage{overpic}
\usepackage{float}  
\usepackage{subfloat}

\usepackage{bm}
\usepackage{multirow}
\usepackage{graphicx}
\usepackage{subcaption}
\usepackage{colortbl}
\usepackage{enumitem}
\usepackage{stfloats} 
\usepackage{url}
\usepackage{verbatim}
\usepackage[normalem]{ulem}
\useunder{\uline}{\ul}{}
\usepackage[skip=0.5\baselineskip]{caption}
\usepackage{color, xspace}
\usepackage{tcolorbox}  
\tcbuselibrary{breakable}   
\usepackage{pifont}
\usepackage{lipsum}
\usepackage{tcolorbox}

\usepackage{xspace}

\usepackage{marvosym}
\usepackage{ifsym}

\copyrightyear{2025}
\acmYear{2025}
\setcopyright{acmlicensed}\acmConference[CIKM '25]{Proceedings of the 34th ACM International Conference on Information and Knowledge Management}{November 10--14, 2025}{Seoul, Republic of Korea}
\acmBooktitle{Proceedings of the 34th ACM International Conference on Information and Knowledge Management (CIKM '25), November 10--14, 2025, Seoul, Republic of Korea}
\acmDOI{10.1145/3746252.3761427}
\acmISBN{979-8-4007-2040-6/2025/11}
\author{Tongzhou Wu}
\affiliation{%
  \institution{City University of Hong Kong}
  \city{Hong Kong}
  \country{China}}
\email{tongzhowu3-c@my.cityu.edu.hk}

\author{Yuhao Wang}
\affiliation{%
  \institution{City University of Hong Kong}
  \city{Hong Kong}
  \country{China}}
\email{yhwang25-c@my.cityu.edu.hk}

\author{Maolin Wang}
\affiliation{%
 \institution{City University of Hong Kong}
 \city{Hong Kong}
 \country{China}}
\email{Morin.wang@my.cityu.edu.hk}

\author{Chi Zhang}
\affiliation{%
  \institution{Harbin Engineering University}
  \city{Harbin}
  \country{China}}
\email{zhangchi20@hrbeu.edu.cn}

\author{Xiangyu Zhao\Letter}
\thanks{\Letter \text{Corresponding author}}
\affiliation{%
 \institution{City University of Hong Kong}
 \city{Hong Kong}
 \country{China}}
\email{xianzhao@cityu.edu.hk}

\renewcommand{\shortauthors}{Tongzhou Wu et al.}

\begin{document}
\title{Empowering Denoising Sequential Recommendation with Large Language Model Embeddings}

\begin{abstract}
Sequential recommendation aims to capture user preferences by modeling sequential patterns in user-item interactions. However, these models are often influenced by noise such as accidental interactions, leading to suboptimal performance. Therefore, to reduce the effect of noise, some 
works propose explicitly identifying and removing noisy items. 
However, we find that simply relying on collaborative information may result in an over-denoising problem, especially for cold items.  
To overcome these limitations, we propose a novel framework: Interest Alignment for Denoising Sequential Recommendation (IADSR) which integrates both collaborative and semantic information. 
Specifically, IADSR is comprised of two stages: in the first stage, we obtain the collaborative and semantic embeddings of each item from a traditional sequential recommendation model and an LLM, respectively. In the second stage, we align the collaborative and semantic embeddings and then identify noise in the interaction sequence based on long-term and short-term interests captured in the collaborative and semantic modalities. 
Our extensive experiments on four public datasets validate the effectiveness of the proposed framework and its compatibility with different sequential recommendation systems. The code and data are released for reproducibility: 
\url{https://github.com/Applied-Machine-Learning-Lab/IADSR}.
\end{abstract}
\keywords{Denoising, Sequential Recommendation, Recommender System, Large Language Model, User Interest}
\ccsdesc[500]{Information systems~Recommender systems}

\maketitle
\section{Introduction} \label{sec:intro}
\label{sec:introduction}
 
In recent years, recommender systems have become indispensable components of modern digital platforms, serving as essential tools to alleviate information overload and enhance user experience across diverse domains such as news~\cite{raza2022news}, entertainment services~\cite{aggarwal2019recommendation} and social media~\cite{zhou2019online}. Among various recommendation tasks, sequential recommendation has gained considerable attention since it aims to capture the temporal dynamics of user behavior and sequential dependencies in interaction histories~\cite{zhao2018deep,fang2020deep,liu2023exploration}, enabling more accurate and dynamic predictions of future interactions~\cite{ye2020time}.
Recent advances in denoising sequential recommendation have explored diverse neural architectures, including Recurrent Neural Networks (RNNs)~\cite{choe2021recommendation,jiang2017play}, convolutional neural network (CNNs)~\cite{tang2018personalized,yan2019cosrec}, Graph Neural Networks (GNNs)~\cite{fan2023graph,zhang2024ssdrec}, and Transformer-based models~\cite{kang2018self,sun2019bert4rec,li2023strec,liu2024large}. However, these approaches primarily focus on improving recommendation performance through architectural modifications, without explicitly addressing the inherent noise present in interaction sequences.

Therefore, despite remarkable progress, sequential recommendation faces significant challenges in real-world applications due to noisy interactions~\cite{wang2019sequential}. Specifically, Such noise includes accidental clicks~\cite{zhang2023denoising}, exploratory behaviors, or interactions that do not reflect true user preferences~\cite{jain2023sampling,o2006detecting,zhang2021fscr}. This problem is exacerbated in sequential scenarios as noise propagates through the modeling process~\cite{chen2022denoising,han2024end4rec,gao2024smlp4rec}, potentially leading to misinterpreted user intentions~\cite{zhang2024ssdrec,chen2023bias}. The presence of noisy interactions could severely distort the learned user preferences and sequential patterns, ultimately degrading recommendation quality~\cite{martinez2016managing,jeong2022fpadametric, lin2023self}.

To address this challenge, denoising sequential recommendation has emerged as a promising research direction~\cite{li2021debiasing}. Early approaches primarily focused on identifying and filtering noisy interactions based on collaborative signals derived from user-item interaction matrices~\cite{li2021discovering,wang2024collaborative,cheng2024empowering} (e.g., directly removing identified noisy interactions~\cite{wang2021denoising,yang2023debiased}, or replace noisy items with alternative items that better align with the user's established preference patterns~\cite{lin2023self,zhang2022hierarchical,zhang2024ssdrec}). These tasks leverage patterns in collective user behaviors to distinguish real preferences from noise and aim to capture better user preferences and behavioral patterns embedded within historical interaction sequences, thereby enhancing recommendation accuracy and relevance~\cite{zhang2019deep}. 
However, relying solely on collaborative information presents inherent limitations, particularly for cold items with sparse interaction histories ~\cite{schein2002methods,yang2018understanding,zhao2003collaborative}. Without leveraging content features, these models lack contextual understanding of interactions, making it difficult to differentiate between real user preferences and random behaviors~\cite{jhamb2018attentive,chen2025dual,fu2025unified}. Consequently, it could lead to over-denoising issues, i.e., a large proportion of cold items would be identified as noise and removed, which potentially dampens the denoising performance.



\begin{figure}[t]
\small
\setlength\abovecaptionskip{0.1\baselineskip}
\setlength\belowcaptionskip{-1.5\baselineskip}
    \centering
    \begin{minipage}{0.285\linewidth}
        \centering
        \begin{subfigure}{1\linewidth}
        \includegraphics[width=0.995\linewidth]{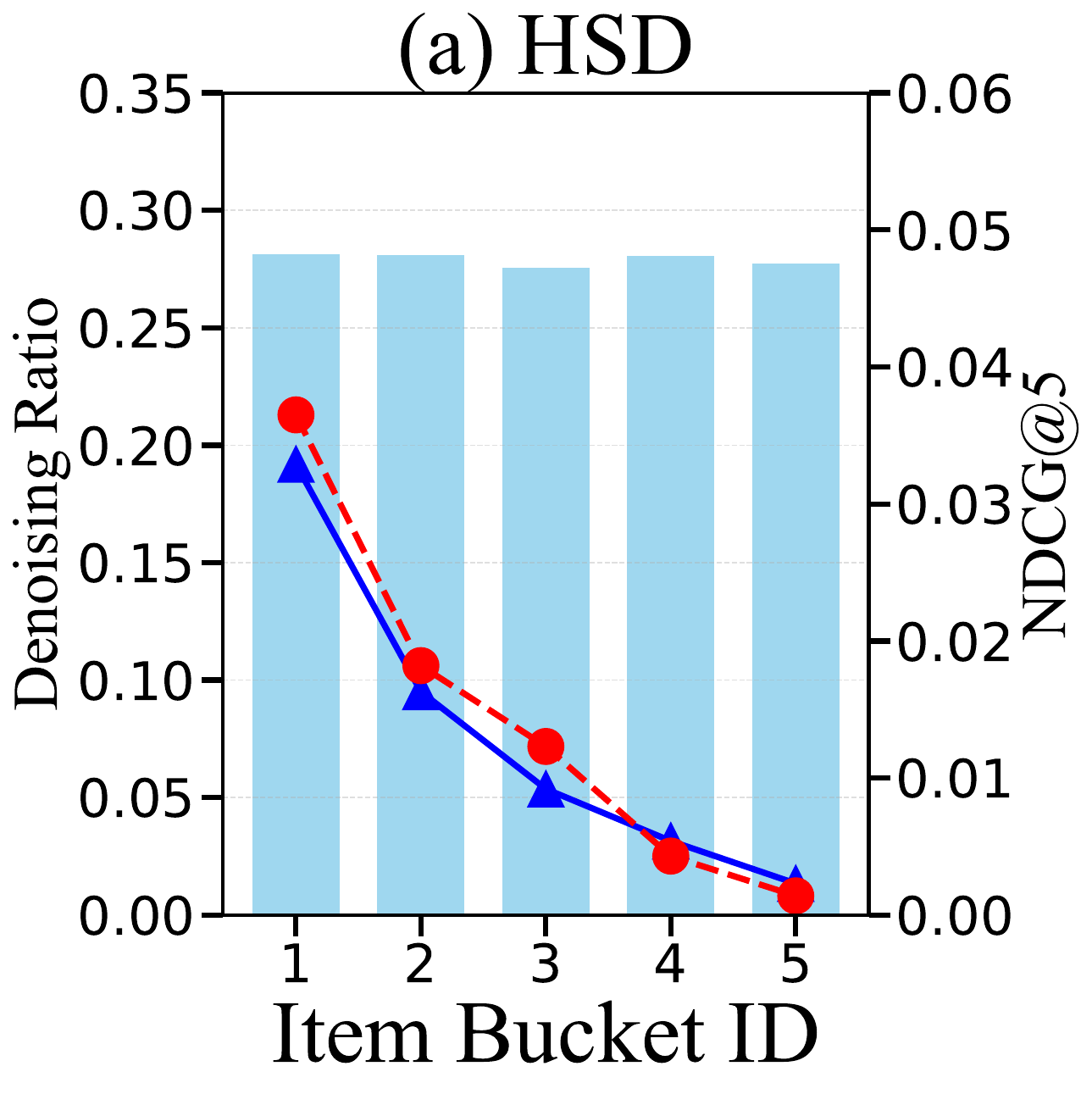}
        \vspace{0mm}
        \label{fig:RQ2-1}
        \end{subfigure}
    \end{minipage}
    \begin{minipage}{0.285\linewidth}
        \centering
        \begin{subfigure}{1\linewidth}
        \includegraphics[width=0.995\linewidth]{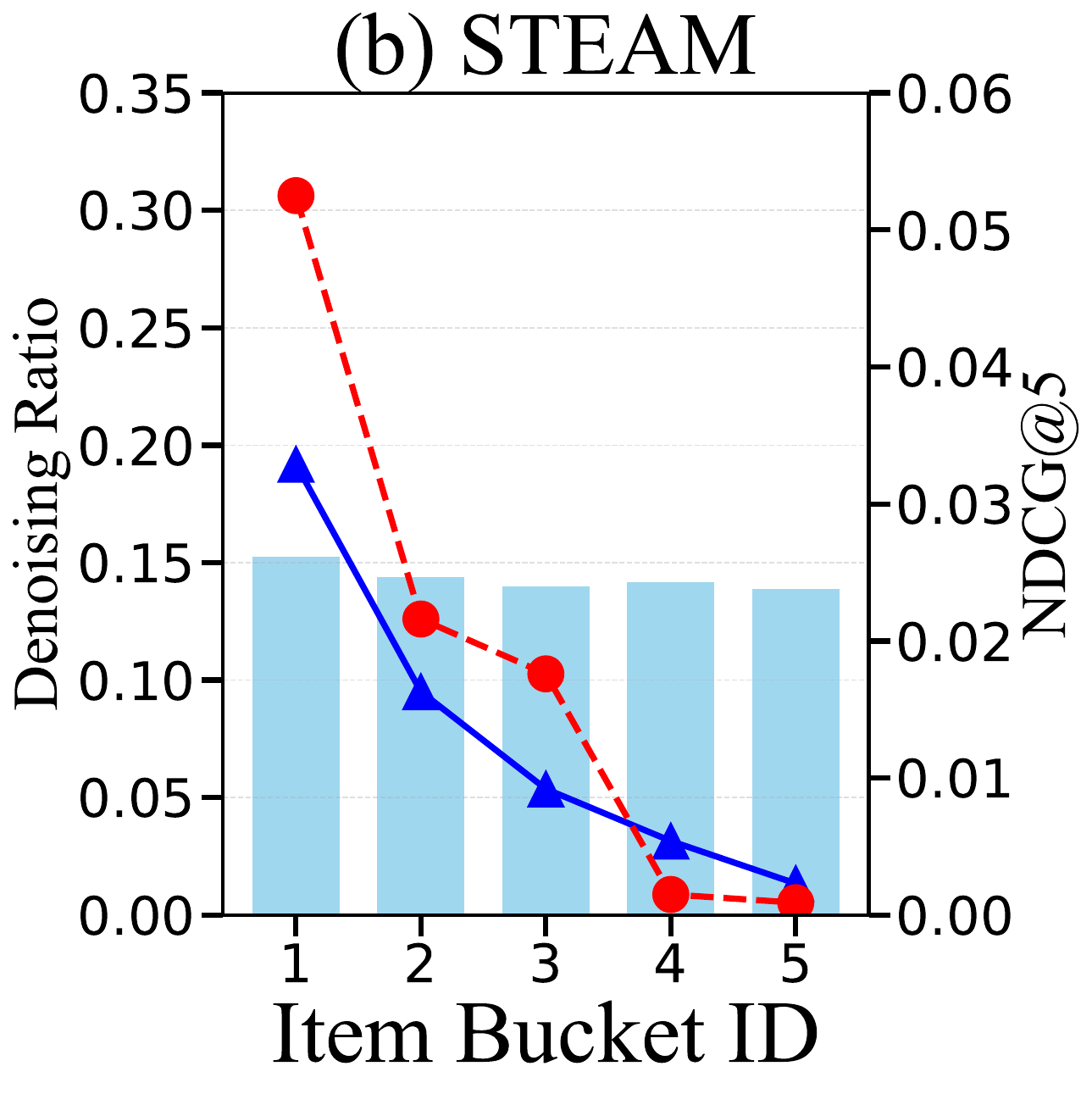}
        \vspace{0mm}
        \label{fig:RQ2-2}
        \end{subfigure}
    \end{minipage}
    \begin{minipage}{0.40\linewidth}
        \centering
        \begin{subfigure}{1\linewidth}
        \includegraphics[width=0.995\linewidth]{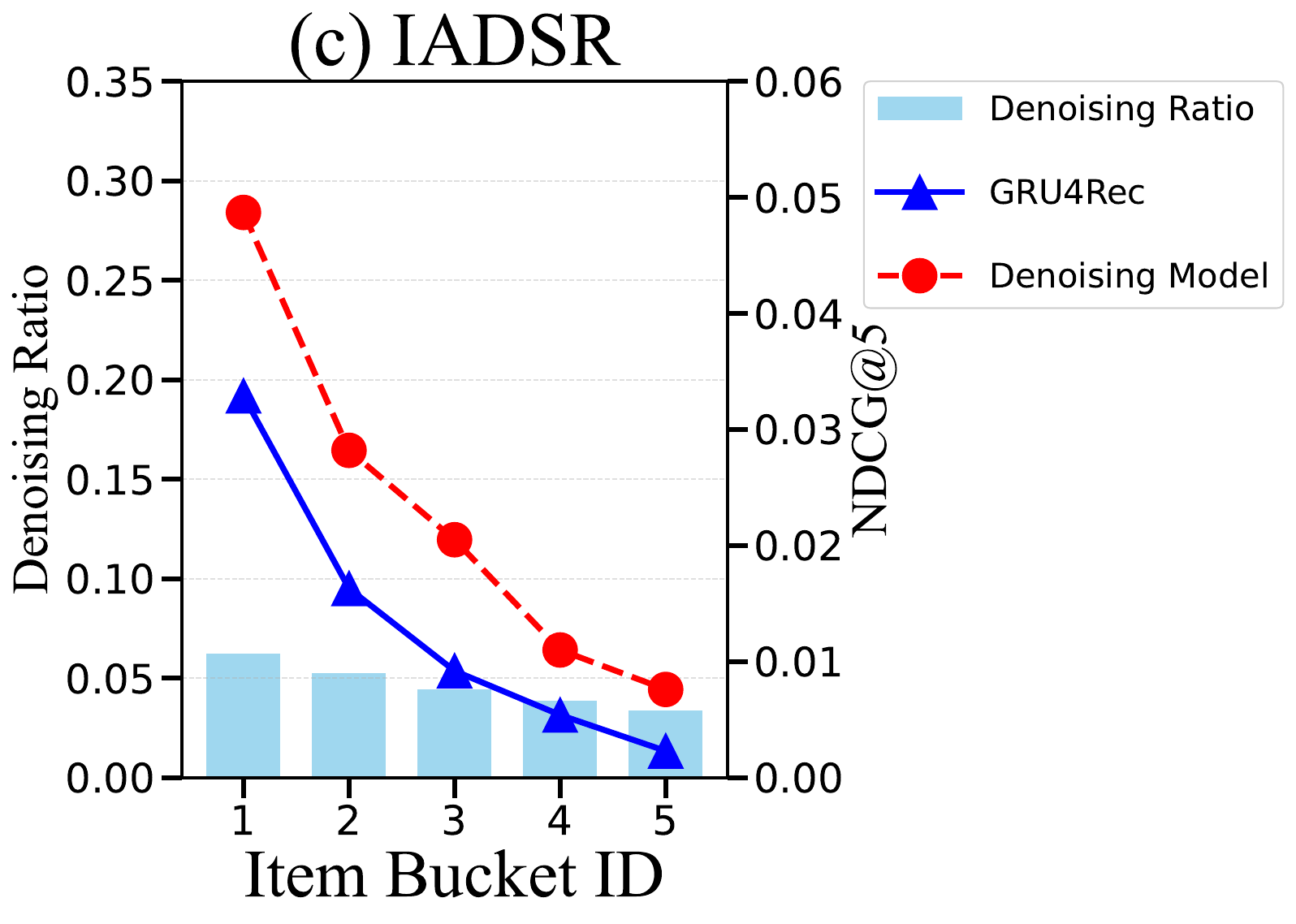}
        \vspace{0mm}
        \label{fig:RQ2-3}
        \end{subfigure}
    \end{minipage}
    \vspace{-4mm}
    \caption{Performance of HSD, STEAM, and our proposed IADSR on Beauty dataset. Item bucket ID from 1 to 5 denotes hot to cold. The denoising ratio represents the percentage of items removed as noise, and the recommendation model used for comparison is GRU4Rec.} 
    \label{fig:RQ2}
\end{figure}
To illustrate the potential over-denoising, we conducted a preliminary analysis of the denoising ratio and accuracy across all the items. The denoising ratio refers to the proportion of items in the original interaction sequence that are identified as noise by each method. As shown in Figure 1, we compared the denoising ratios and recommendation performance of HSD~\cite{zhang2022hierarchical}, STEAM~\cite{lin2023self}, and IADSR on the Beauty dataset. We divided the original sequence into five equal-frequency buckets based on item popularity (from hot to cold items) and then evaluated the recommendation accuracy using NDCG@5 on each bucket. The blue bars represent the denoising ratio applied by each method, while the line graphs illustrate the recommendation performance. The blue curve represents the results tested with the GRU4Rec model.
We can observe that across all methods, as item popularity decreases from bucket 1 (hottest) to bucket 5 (coldest), recommendation performance significantly declines. This uniform filtering process fails to account for the unique characteristics of cold items, potentially removing interactions that may seem unreasonable from a purely collaborative perspective but could actually reflect real user interests in less popular items~\cite{kong2019classical,zhang2024ssdrec,chen2014marginalized}. However, IADSR performs better than baseline approaches, particularly for cold items. While the absolute improvement may appear larger on popular items due to their higher baseline performance, the relative improvement of IADSR is actually more pronounced for cold items. By preserving more useful information during the denoising process for these less popular items, IADSR achieves consistently better recommendation accuracy across all buckets, with the advantage becoming more pronounced for the coldest items in buckets 4 and 5. 

Therefore, relying solely on collaborative signals may be insufficient~\cite{zhao2018recommendations,rabiah2024bridging,wang2024llm4dsr}, and we consider leveraging the textual information of items to address this limitation. Large Language Models (LLMs), which have gained tremendous popularity in recent years, can effectively complement this aspect~\cite{acharya2023llm,lyu2023llm,zhao2024llm}. We can utilize LLMs to generate embeddings of textual information and combine them with traditional collaborative information for denoising~\cite{al2015semantic}. By aligning embeddings from both modalities, we attempt to better capture users' genuine preferences. 

Consequently, we present IADSR, a novel two-stage denoising framework that effectively integrates semantic information from LLMs with collaborative signals for enhanced denoising. Our framework operates through two distinct stages: (1) dual representation learning, where we independently obtain item embeddings from both LLMs and traditional sequential models; (2) cross-modal alignment and noise identification, where we leverage long-term and short-term user interests to detect and filter noisy interactions.

The main contributions of this paper are as follows.

\begin{itemize}[leftmargin=*]
	\item We propose IADSR, a novel denoising paradigm for sequential recommendation compatible with diverse backbone models and achieving performance enhancement.
	\item The proposed framework combines LLMs with sequential recommendation without fine-tuning.
	\item Experiments on four public datasets, i.e., Amazon Beauty, Sports, Toys, and MovieLens-100K, have demonstrated the effectiveness of our proposed method.
\end{itemize}

\section{Preliminary} \label{sec:preliminary}
In this section, we introduce the basic notations and definitions used throughout this work.
\subsection{Sequential Recommendation}

Let $\mathcal{U} = \{u_1, u_2, ..., u_m\}$ denotes the set of users and $\mathcal{I} = \{i_1, i_2, ..., i_n\}$ denotes the set of items. For each user $u \in \mathcal{U}$ 
its interaction sequence $S^u = [i^u_1, i^u_2, ..., i^u_t]$ is sorted in ascending order by time, where $i^u_j \in \mathcal{I}$ represents the $j$-th item that user $u$ has interacted with. The goal of sequential recommendation is to predict the next item $i^u_{t+1}$ that user $u$ is likely to interact with based on $S^u$.

Formally, let $\hat{\mathbf{y}}^u \in \mathbb{R}^{|I|}$ denote the prediction scores for all items in the item set $I$, where each element $\hat{y}^u_i$ represents the predicted probability that user $u$ will interact with item $i$ next. The ground truth is typically represented as a one-hot encoded vector $\mathbf{y}^u$, where $y^u_i = 1$ if item $i$ is the actual next item in the sequence ($i = i^u_{t+1}$), and $y^u_i = 0$ otherwise.

The Cross-Entropy loss function is then defined as:
\begin{equation}
\mathcal{L}_{CE} = -\sum_{u \in \mathcal{U}} \sum_{i \in \mathcal{I}} y^u_i \log(\hat{y}^u_i)
\end{equation}

This loss function encourages the model to assign a high probability to the correct next item. However, in the presence of noisy interactions in the sequence, optimizing solely based on this loss can lead the model to learn patterns from noise, potentially degrading recommendation performance.

\subsection{Denoising Sequential Recommendation}


In the denoising sequential recommendation task, for a user $u$ with interaction sequence $S^u = [i^u_1, i^u_2, ..., i^u_t]$, we need to identify and remove the noise interactions:
\begin{equation}
S^u_{denoised} = S^u \setminus S^u_{noise}
\end{equation}

\noindent where $S^u_{denoised}$ contains interactions that truly represent user $u$'s preferences, and $S^u_{noise}$ consists of noisy interactions that may mislead the recommendation model~\cite{steck2013evaluation,wang2024trustworthy}.

\section{Method} \label{sec:method}
In this section, an overview of the proposed framework is first provided, followed by details of different modules.

\subsection{Overview}

In this section, we introduce the overall framework of IADSR, which employs a two-stage framework to enhance recommendation quality by identifying and removing noise from user interaction sequences. As depicted in Figure 1, in the first stage we construct item embeddings from two distinct sources by processing raw user sequences through parallel paths: (1) extracting semantic representations via LLM encoding of textual descriptions, generating comprehensive semantic understanding~\cite{hu2024enhancing,liu2024llm,harte2023leveraging}; (2) learning collaborative patterns through traditional sequential models using item ID representations. These complementary embeddings capture both content semantics and user behavior patterns~\cite{zhang2022latent,liu2017multi}. In the second stage, we align the collaborative and semantic embeddings through cosine similarity measures to identify noise in user sequences~\cite{zhang2024m3oe}. We compute similarity scores between long-term and short-term interest representations from both embedding spaces, then apply a Gumbel-Sigmoid function to generate binary masks indicating noisy items.


\begin{figure*}
\setlength\abovecaptionskip{0.1\baselineskip}
\setlength\belowcaptionskip{0.2\baselineskip}
\centering
    
	\includegraphics[width=0.96\linewidth]{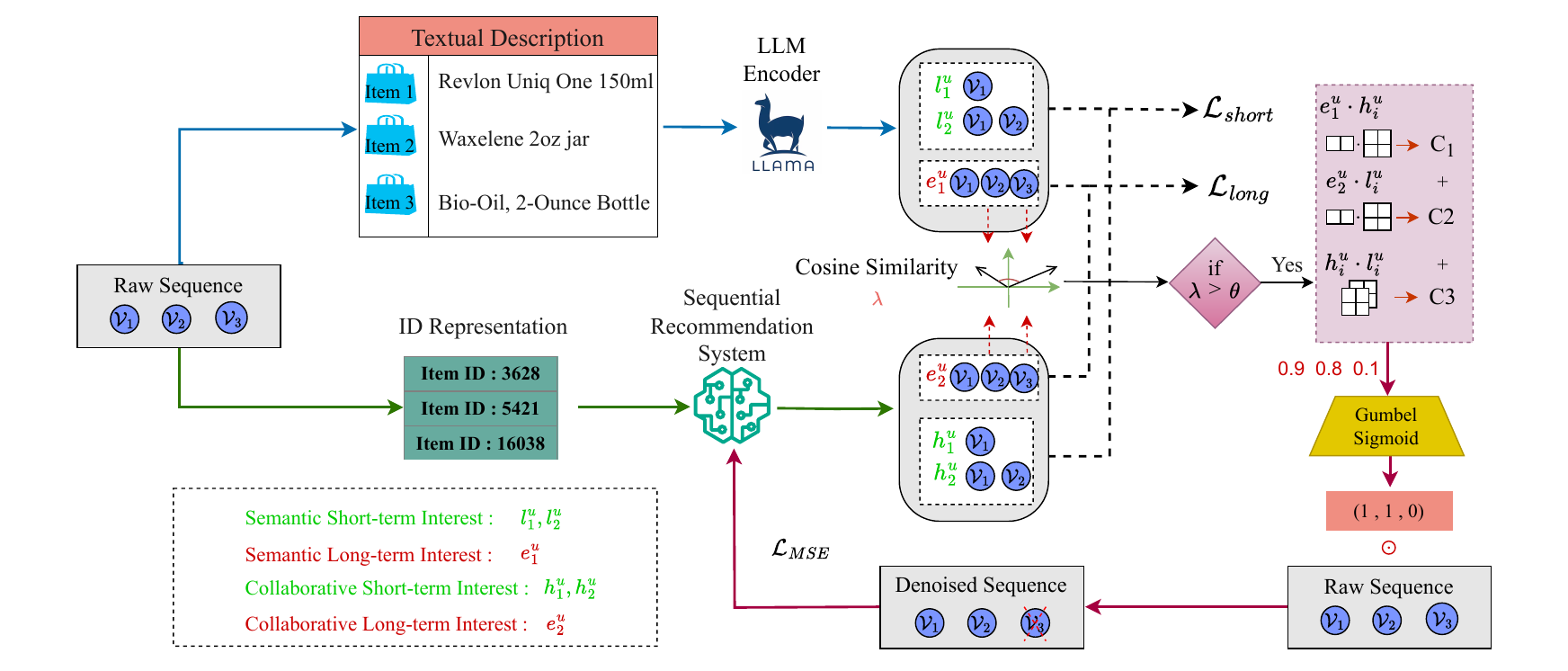}
	\caption{Overview of the IADSR framework. The black arrow denotes the data flow. The blue and green color denotes semantic and collaborative modality, respectively.}
	\label{fig:Fig1_DRS}
	\vspace*{-4mm}
\end{figure*}

\subsection{{Semantic Encoding}}

Previous studies have not focused on extracting semantic information, possibly due to high computational costs and insufficient world knowledge. Recently, Large Language Models (LLMs) have demonstrated remarkable capabilities in understanding text across diverse domains~\cite{li2023hamur}. However, these decoder-only architectures exhibit limitations in encoding capabilities, as they are optimized for generation rather than representation~\cite{boz2024improving}. In recommender systems, while LLMs can extract semantic information from text, their architectural constraints may result in suboptimal embeddings compared to dedicated encoding frameworks~\cite{li2023prompt}.

For our approach, we employ LLM2Vec~\cite{behnamghader2024llm2vec} to generate semantic embeddings for each item. Specifically, LLM2Vec addresses the encoding limitations of decoder-only Large Language Models by efficiently extracting high-quality semantic embeddings that capture nuanced user preferences from textual descriptions, enhancing recommender systems with encoder capabilities.
This enables the extraction of high-quality semantic embeddings without additional fine-tuning, making it ideal for our recommendation scenario.  

For each item $i \in \mathcal{I}$, we obtain its item name $Z_i$ and process it through LLM2Vec to generate semantic embeddings:
\begin{equation}
\mathbf{e}^{LLM}_i = \text{LLM2Vec}(Z_i)
\end{equation}
\noindent where $\mathbf{e}^{LLM}_i \in \mathbb{R}^d$ is the semantic embedding vector of item $i$ with dimension $d$, and $\text{LLM2Vec}(\cdot)$ represents the encoding function that maps item names to dense vector embeddings.

These LLM-integrated embeddings capture semantic relationships between items based on their names, offering rich complementary information to collaborative signals. Even with just the item name, the semantic embeddings effectively capture product categories and attributes through the pre-trained knowledge embedded in the LLM. This approach is particularly valuable for cold items with limited interaction histories, as it allows the model to better infer item similarities from semantic meaning rather than relying solely on interaction patterns.

\subsection{Interest Alignment}
Despite differences between collaborative and semantic modalities in their representational spaces, we posit that for any given user, their fundamental interests should remain consistent regardless of which modality is used to represent them~\cite{zhao2023kuaisim}. Both modalities ultimately attempt to capture the same underlying user interests.

In order to effectively leverage both semantic and collaborative information for optimal merging, we first organize user interests into long-term and short-term embeddings and then systematically align these interests across modalities.

\subsubsection{Interest Representation}

For each user $u$ with interaction sequence $S^u$ = $\{i_1^u, i_2^u, \ldots, i_n^u\}$, we structure their interests as: 

\textbf{Long-term Interests:} The long-term interest represents a comprehensive view of the user's preferences across their entire interaction history. Rather than encoding each item separately, we encode the complete sequence of user interactions directly:
\begin{equation}
\mathbf{e}_1^u = \text{LLM2Vec}(S^u) 
\end{equation}

\noindent where $\text{LLM2Vec}(S^u)$ processes the entire interaction sequence as a single input, capturing the holistic semantic meaning of the user's complete interaction history.

\textbf{Short-term Interests:} The short-term interests capture the evolving preferences at different time steps. For each time step $t$, we independently encode the partial sequence up to that point:
\begin{equation}
\mathit{l}_t^u = \text{LLM2Vec}(S^u_t), \quad \forall t \in \{1, 2, \ldots, n-1\}
\end{equation}

\noindent where $S^u_t = \{i_1^u, i_2^u, \ldots, i_t^u\}$ is the complete subsequence of interactions up to time step $t$. This results in a total of $n-1$ distinct separate encodings, each representing the user's interests at a different point in their interaction history.

Similarly, for the collaborative interests derived from the sequential recommendation models:
\begin{equation}
\mathbf{e}_2^u = \text{SRS}(S^u)
\end{equation}
\begin{equation}
\mathbf{h}_t^u = \text{SRS}(S^u_t), \quad \forall t \in \{1, 2, \ldots, n-1\}
\end{equation}

\noindent where $\text{SRS}(\cdot)$ represents the encoding function of the sequential recommendation models.

In summary, we extract two types of interest embeddings:
\begin{itemize}[leftmargin=*]
    \item \textbf{Semantic Interests}: Long-term semantic interest $\mathbf{e}_1^u$ and corresponding short-term semantic interests $\{\mathit{l}_1^u, \mathit{l}_2^u, \ldots, \mathit{l}_{n-1}^u\}$ effectively derived from LLM embeddings.
    \item \textbf{Collaborative Interests}: Long-term collaborative interest $\mathbf{e}_2^u$ and short-term collaborative interests $\{\mathbf{h}_1^u, \mathbf{h}_2^u, \ldots, \mathbf{h}_{n-1}^u\}$ derived from the sequential model.
\end{itemize}

\subsubsection{Cross-Modal Interest Alignment}

To effectively combine the strengths of both semantic and collaborative information, we align these interest embeddings using InfoNCE loss~\cite{oord2018representation,he2020momentum}. This alignment maximizes the mutual information between corresponding interest embeddings from different modalities~\cite{rusak2024infonce}.

The InfoNCE loss for interest alignment is formulated as:
\begin{align}
\mathcal{L}_{Info} &= \mathcal{L}_{long} + \mathcal{L}_{short} \\
\mathcal{L}_{long} &= -\frac{1}{N}\sum_{i=1}^{N} \log \frac{e^{sim(\mathbf{e}_2^i, \mathbf{e}_1^i)/\tau}}{\sum_{j=1}^{N} e^{sim(\mathbf{e}_2^i, \mathbf{e}_1^j)/\tau}} \\
\mathcal{L}_{short} &= -\frac{1}{M}\sum_{i=1}^{M} \log \frac{e^{sim(\mathbf{h}_i, \mathit{l}_i)/\tau}}{\sum_{j=1}^{M} e^{sim(\mathbf{h}_i, \mathit{l}_j)/\tau}}
\end{align}

\noindent where $N$ denotes the batch size, representing the number of users processed in each training iteration and $M$ represents the total number of short-term interest embeddings in the batch, $sim(\cdot,\cdot)$ is the cosine similarity function and $\tau$ is a temperature parameter. Specifically, $\mathcal{L}_{long}$ represents the alignment loss between long-term semantic interest $\mathbf{e}_1^u$ and long-term collaborative interest $\mathbf{e}_2^u$, while $\mathcal{L}_{short}$ represents the alignment loss between short-term semantic interests $\{\mathit{l}_1^u, \mathit{l}_2^u, \ldots, \mathit{l}_{n-1}^u\}$ and short-term collaborative interests $\{\mathbf{h}_1^u, \mathbf{h}_2^u, \ldots, \mathbf{h}_{n-1}^u\}$.

Through this alignment process, IADSR creates unified interest embeddings that leverage both the semantic understanding from LLMs and the collaborative patterns from sequential models, providing a more comprehensive basis for subsequent noise detection.

\subsection{Sequence Denoising}

After aligning interest embeddings across modalities, we proceed to identify and then filter out noise in user interaction sequences. This is particularly difficult since noise is often contextual and user-dependent without explicit labels. IADSR leverages the complementary nature of semantic and collaborative embeddings to distinguish real user preferences from noise that the traditional single-modal approaches might miss.

For each user $u$, we compute the cosine similarity between their long-term collaborative interest $\mathbf{e}_2^u$ and the corresponding long-term semantic interest $\mathbf{e}_1^u$:
\begin{equation}
\text{sim}_{long}(u) = \cos(\mathbf{e}_1^u, \mathbf{e}_2^u)
\end{equation}

Users whose cross-modal similarity $\lambda$ exceeds a threshold $\theta$ proceed to the detailed noise detection stage:
\begin{equation}
\text{qualified}(u) = \mathbf{1}[\text{sim}_{long}(u) \geq \theta]
\end{equation}
where $\mathbf{1}[\cdot]$ is the indicator function. This preliminary filtering ensures that we only apply denoising to users whose embeddings show sufficient cross-modal consistency, avoiding potentially harmful modifications to sequences where modality alignment is poor.

\subsubsection{Item-level Noise Detection}

For qualified users, we perform item-level noise detection by examining the consistency between different interest embeddings. For each time step $t$ in the qualified user's sequence, we compute three similarity scores between the corresponding interest embeddings as  $\left(c_1, c_2, c_3\right)$:
\begin{align}
c_1(t) &= \cos(\mathbf{e}_1^u, \mathbf{h}_t^u) \\
c_2(t) &= \cos(\mathbf{e}_2^u, \mathit{l}_t^u) \\
c_3(t) &= \cos(\mathbf{h}_t^u, \mathit{l}_t^u)
\end{align}

These scores respectively measure: (1) the semantic long-term to collaborative short-term consistency, (2) the collaborative long-term to semantic short-term consistency, and (3) the overall short-term cross-modal consistency. This design ensures that potential noise can be detected from complementary perspectives across modalities and interest levels. We combine these scores to obtain a comprehensive noise indicator:
\begin{equation}
\text{score}(t) = c_1(t) + c_2(t) + c_3(t)
\end{equation}

\subsubsection{Mask Generation via Gumbel-Sigmoid}

To convert the continuous noise scores into binary denoising decisions, we employ the robust Gumbel-Sigmoid function~\cite{liu2021gumbel}. The Gumbel-Sigmoid function enables differentiable binary sampling during training by adding Gumbel noise to logits and applying a temperature-controlled sigmoid function, allowing models to make discrete decisions (like masking noise) while effectively maintaining gradient flow for end-to-end training~\cite{jang2016categorical,ng2022masked}:
\begin{align}
m_t &= \text{GumbelSigmoid}(\text{score}(t), \tau, \text{hard}=\text{True}) \notag\\
&= \begin{cases}
\mathbf{1}[y_t > 0.5], & \text{if hard}=\text{True} \\
y_t, & \text{if hard}=\text{False} \\
\end{cases}
\end{align}
\begin{align}
y_t &= \delta\left(\frac{\text{score}(t) + g_t}{\tau}\right) \notag\\
g_t &= -\log(-\log(U_t + \epsilon) + \epsilon) \notag\\
U_t &\sim \text{Uniform}(0, 1) \notag
\end{align}
Here, $\tau$ is the temperature parameter controlling the smoothness of the approximation, $\delta$ is the sigmoid function, $U_t$ is a uniform random variable, and $\epsilon$ is a constant added for numerical stability. When $\text{hard}=\text{True}$, we discretize the output to binary values while preserving gradients through a straight-through estimator.

The resulting mask $m_t \in \{0, 1\}$ indicates whether each interaction should be preserved (1) or filtered out (0) as noise. By incorporating multiple similarity measures and maintaining differentiability, our model can make effective discrete denoising decisions while learning from its own denoising process through backpropagation.

The denoised sequence for user $u$ is then obtained by applying the mask to the original sequence:
\begin{equation}
S^u_{denoised} = \{i_t^u \mid m_t = 1, t = 1,2,\ldots,n\}
\end{equation}

\subsection{Sequence Reconstruction}

To prevent the loss of critical signals from "over-denoising" (especially for cold items), we use a sequence reconstruction mechanism to balance noise removal with the preservation of user preferences.

\subsubsection{Progressive Denoising Process}

Our approach employs a progressive denoising strategy across training epochs. For each user, we apply the mask learned from the previous epoch to the original sequence embeddings, preserving the model's incremental learning process while always anchoring to the original data. Formally, for a user $u$ at epoch $e$:
\begin{equation}
\mathbf{X}_u^{(e)} = \mathbf{X}_u^{original} \odot \mathbf{K}_u^{(e-1)}
\end{equation}

\noindent where $\mathbf{X}_u^{(e)}$ represents the input sequence embedding at epoch $e$, $\mathbf{X}_u^{original}$ is the original unmodified sequence embedding, $\mathbf{K}_u^{(e-1)}$ is the binary mask generated from the previous epoch, and $\odot$ denotes element-wise multiplication. For the initial epoch ($e=0$), we use the original sequence without masking.

\subsubsection{Decoder-based Reconstruction}

To ensure that the denoising process preserves essential information, we employ a decoder to reconstruct the original sequence from the denoised representation:
\begin{equation}
\hat{\mathbf{X}}_u = \text{Decoder}(\mathbf{H}_u \odot \mathbf{K}_u)
\end{equation}

\noindent where $\hat{\mathbf{X}}_u$ represents the reconstructed sequence embedding for user $u$, $\mathbf{H}_u$ represents the model's hidden states (i.e., GRU output embeddings), $\mathbf{K}_u$ is the dynamically generated binary mask for the current epoch, and $\odot$ denotes element-wise multiplication. This process effectively transforms the denoised hidden states back into the original embedding space, allowing us to directly compare the reconstruction with the original input embeddings.

\subsubsection{Reconstruction Loss}

To ensure our denoising process preserves real user preferences while removing only truly noisy interactions, we introduce a reconstruction objective that:
\begin{itemize}[leftmargin=*]
    \item Encourages selective denoising by penalizing the removal of real preference signals.
    \item Provides additional training supervision that helps the model learn more robust embeddings.
    \item Anchors the denoised embeddings to the original data, preventing representation drift.
\end{itemize}

We implement this objective by systematically minimizing the mean squared error between the reconstructed sequence and the corresponding original sequence embeddings:
\begin{equation}
\mathcal{L}_{recon} = \frac{1}{|\mathcal{U}_{mask}|} \sum_{u \in \mathcal{U}_{mask}} ||\hat{\mathbf{X}}_{u} - \mathbf{X}_{u}^{original}||_2^2
\end{equation}

\noindent where $\mathcal{U}_{mask}$ represents the set of users who passed the initial cross-modal consistency check. This squared L2 distance effectively captures the overall reconstruction quality across all dimensions of the embedding space.

The total loss for our model combines the three components:
\begin{equation}
\mathcal{L}_{total} = \mathcal{L}_{CE} + \mathcal{L}_{Info} + \mathcal{L}_{recon}
\end{equation}

\noindent where $\mathcal{L}_{CE}$ is the standard cross-entropy loss for next item prediction, $\mathcal{L}_{Info}$ is the interest alignment loss, and $\mathcal{L}_{recon}$ is the sequence reconstruction loss.
\section{Experiments} \label{sec:experiments}

In this section, we present the experiment results on four public datasets to validate the effectiveness of our methods. Our evaluation is guided by the following research questions:

\begin{itemize}[leftmargin=*]
    \item \textbf{RQ1:} How does IADSR perform compared with the state-of-the-art denoising baseline methods?
        
    \item \textbf{RQ2:} Is IADSR compatible with different sequential recommendation models?
        
    \item \textbf{RQ3:} What impact do the proposed loss functions have on the recommendation performance in IADSR?

    \item \textbf{RQ4:} How sensitive is IADSR to hyperparameters?

    \item \textbf{RQ5:} How do the introduced semantic embeddings contribute to denoising in IADSR?
\end{itemize}


\begin{table}[t]
\centering
\small
\setlength\abovecaptionskip{0\baselineskip}
\setlength\belowcaptionskip{0\baselineskip}
\caption{Experimental data statistics.}
\label{tab:datasets}
\begin{tabular}{lccccc}
\toprule
\textbf{Dataset} & \textbf{\# Users} & \textbf{\# Items} & \textbf{\# Actions} & \textbf{Avg. len} & \textbf{Sparsity} \\
\midrule
Beauty & 22,363 & 12,101 & 198,502 & 8.9 & 99.93\% \\
Sports & 35,598 & 18,357 & 296,337 & 8.3 & 99.95\% \\
Toys & 19,412 & 11,924 & 167,597 & 8.6 & 99.93\% \\
ML-100K & 943 & 1,682 & 100,000 & 106.0 & 93.70\% \\
\bottomrule
\vspace{-7mm}
\end{tabular}
\end{table}

\subsection{Experimental Setting}
\subsubsection{Datasets and Pre-processing}
We conduct experiments on three domains of Amazon datasets and the MovieLens-100K dataset. Their statistics are summarized in Table 1. Average length (Avg. len) represents the mean number of interactions per user, reflecting the length of user behavior sequences, ranging from approximately 8-9 interactions for Beauty, Sports, and Toys datasets to 106 interactions for ML-100K. Sparsity indicates the gap between actual user-item interactions and the theoretically maximum possible interactions in the user-item matrix, demonstrating that all datasets are highly sparse, with Beauty, Sports, and Toys having approximately 99.9\% sparsity, while ML-100K is relatively less sparse at 93.7\%.

\begin{itemize}[leftmargin=*]
   \item \textbf{Amazon}: We utilize three categories from the Amazon review dataset: Beauty, Sports \& Outdoors, and Toys \& Games. For each category, we followed the previous studies~\cite{yuan2021dual,xu2023openp5,barman2024review,zhou2022filter} and adopted the 5-core version where each user and item has at least five interactions, ensuring sufficient sequential patterns for modeling and reducing data sparsity. Each dataset contains user-item interaction records with timestamps, allowing us to construct highly meaningful temporal behavioral sequences. The product metadata in these datasets enables the extraction of semantic information through LLMs.

   \item \textbf{MovieLens-100K}: A widely-used benchmark dataset in the movie recommendation domain, containing 100,000 ratings from users on different movies. It offers a complementary domain to e-commerce and features more structured item attributes.
\end{itemize}

For pre-processing, we follow the standard practices in sequential recommendation as ~\cite{sun2019bert4rec,yuan2021dual,zhang2024ssdrec}. Specifically, each user's interaction sequence is sorted chronologically by timestamp to preserve the temporal order of user behaviors. Based on the observed average sequence lengths in the Amazon datasets, we set the maximum sequence length to 32 for these datasets, while for MovieLens we use a maximum length of 50 to ensure optimal performance.

\begin{table*}[ht]
\centering
\tiny  
\setlength\abovecaptionskip{0\baselineskip}
\setlength\belowcaptionskip{-0.5\baselineskip}
\caption{Overall performance comparison on Beauty, Sports and Toys dataset. Boldface denotes the best result and underline indicates the second-best results. `\textbf{{*}}' denotes significant improvement (i.e., two-sided t-test with p < 0.05).}
\label{tab:overall}
\resizebox{0.77\textwidth}{!}{
\begin{tabular}{c|cc|cccccc}
\toprule
Dataset & \multicolumn{2}{c|}{Model} & HR@5 & HR@10 & HR@20 & NDCG@5 & NDCG@10 & NDCG@20 \\
\midrule
\multirow{14}{*}{Beauty} 
& \multirow{4}{*}{GRU4Rec} & base & 0.0174 & 0.0249 & 0.0351 & 0.0120 & 0.0144 & 0.0169 \\
&  & HSD & \underline{0.0229} & \underline{0.0365} & \textbf{0.0520*} & \underline{0.0141} & \underline{0.0185} & \underline{0.0234} \\
&  & SSDRec & 0.0218 & 0.0360 & \underline{0.0510} & 0.0136 & 0.0181 & 0.0229 \\
&  & IADSR & \textbf{0.0300*} & \textbf{0.0396*} & 0.0486 & \textbf{0.0213*} & \textbf{0.0244*} & \textbf{0.0267*} \\
\cmidrule{2-9}
& \multirow{4}{*}{Caser} & base & 0.0075 & 0.0130 & 0.0216 & 0.0043 & 0.0060 & 0.0082 \\
&  & HSD & 0.0137 & 0.0214 & 0.0327 & \underline{0.0086} & 0.0111 & 0.0140 \\
&  & SSDRec & \underline{0.0142} & \underline{0.0235} & \underline{0.0363} & 0.0085 & \underline{0.0115} & \underline{0.0147} \\
&  & IADSR & \textbf{0.0178*} & \textbf{0.0297*} & \textbf{0.0469*} & \textbf{0.0094*} & \textbf{0.0137*} & \textbf{0.0188*} \\
\cmidrule{2-9}
& \multirow{4}{*}{SASRec} & base & 0.0267 & 0.0385 & 0.0554 & 0.0184 & 0.0218 & 0.0261 \\
&  & HSD & 0.0245 & 0.0436 & 0.0668 & 0.0140 & 0.0201 & 0.0260 \\
&  & SSDRec & \textbf{0.0342*} & \textbf{0.0538*} & \underline{0.0782} & \underline{0.0196} & \underline{0.0259} & \underline{0.0321} \\
&  & IADSR & \underline{0.0323} & \textbf{0.0538*} & \textbf{0.0836*} & \textbf{0.0202*} & \textbf{0.0272*} & \textbf{0.0349*} \\
\cmidrule{2-9}
& \multicolumn{2}{c|}{STEAM} & 0.0292 & 0.0395 & 0.0515 & 0.0201 & 0.0234 & 0.0264 \\
& \multicolumn{2}{c|}{DCRec} & 0.0110 & 0.0202 & 0.0370 & 0.0066 & 0.0095 & 0.0137 \\
\midrule
\multirow{14}{*}{Sports} 
& \multirow{4}{*}{GRU4Rec} & base & 0.0059 & 0.0940 & 0.0150 & 0.0037 & 0.0048 & 0.0060 \\
&  & HSD & \underline{0.0136} & 0.0186 & 0.0303 & 0.0077 & 0.0099 & 0.0129 \\
&  & SSDRec & 0.0122 & \underline{0.0196} & \underline{0.0318} & \underline{0.0083} & \underline{0.0107} & \underline{0.0137} \\
&  & IADSR & \textbf{0.0155*} & \textbf{0.0240*} & \textbf{0.0340*} & \textbf{0.0095*} & \textbf{0.0122*} & \textbf{0.0148*} \\
\cmidrule{2-9}
& \multirow{4}{*}{Caser} & base & 0.0059 & 0.0081 & 0.0118 & 0.0026 & 0.0031 & 0.0048 \\
&  & HSD & \underline{0.0063} & 0.0119 & 0.0212 & \textbf{0.0043*} & 0.0061 & \underline{0.0084} \\
&  & SSDRec & 0.0060 & \underline{0.0123} & \underline{0.0213} & \underline{0.0042} & \underline{0.0062} & \underline{0.0084} \\
&  & IADSR & \textbf{0.0080*} & \textbf{0.0152*} & \textbf{0.0258*} & 0.0041 & \textbf{0.0063*} & \textbf{0.0089*} \\
\cmidrule{2-9}
& \multirow{4}{*}{SASRec} & base & 0.0112 & 0.0178 & 0.0269 & 0.0074 & 0.0093 & 0.0102 \\
&  & HSD & 0.0119 & 0.0202 & 0.0309 & 0.0078 & 0.0108 & 0.0127 \\
&  & SSDRec & \underline{0.0132} & \underline{0.0212} & \underline{0.0338} & \underline{0.0099} & \underline{0.0111} & \underline{0.0131} \\
&  & IADSR & \textbf{0.0155*} & \textbf{0.0260*} & \textbf{0.0383*} & \textbf{0.0101*} & \textbf{0.0114*} & \textbf{0.0139*} \\
\cmidrule{2-9}
& \multicolumn{2}{c|}{STEAM} & 0.0149 & 0.0182 & 0.0250 & 0.0078 & 0.0101 & 0.0122 \\
& \multicolumn{2}{c|}{DCRec} & 0.0080 & 0.0141 & 0.0288 & 0.0068 & 0.0088 & 0.0105 \\
\midrule
\multirow{14}{*}{Toys} 
& \multirow{4}{*}{GRU4Rec} & base & 0.0110 & 0.0125 & 0.0133 & 0.0080 & 0.0085 & 0.0091 \\
&  & HSD & \underline{0.0167} & \underline{0.0266} & \textbf{0.0413*} & \underline{0.0109} & \underline{0.0142} & \underline{0.0179} \\
&  & SSDRec & 0.0140 & 0.0227 & 0.0354 & 0.0090 & 0.0118 & 0.0149 \\
&  & IADSR & \textbf{0.0189*} & \textbf{0.0281*} & \underline{0.0389} & \textbf{0.0130*} & \textbf{0.0160*} & \textbf{0.0186*} \\
\cmidrule{2-9}
& \multirow{4}{*}{Caser} & base & 0.0054 & 0.0089 & 0.0145 & 0.0035 & 0.0046 & 0.0060 \\
&  & HSD & \underline{0.0066} & \underline{0.0124} & 0.0192 & 0.0041 & \underline{0.0060} & 0.0076 \\
&  & SSDRec & 0.0065 & 0.0116 & \underline{0.0198} & \underline{0.0044} & \underline{0.0060} & \underline{0.0081} \\
&  & IADSR & \textbf{0.0098*} & \textbf{0.0163*} & \textbf{0.0224*} & \textbf{0.0080*} & \textbf{0.0108*} & \textbf{0.0129*} \\
\cmidrule{2-9}
& \multirow{4}{*}{SASRec} & base & 0.0288 & 0.0394 & 0.0468 & 0.0162 & 0.0216 & 0.0254 \\
&  & HSD & \underline{0.0299} & 0.0451 & 0.0649 & \textbf{0.0180*} & \underline{0.0229} & 0.0279 \\
&  & SSDRec & \textbf{0.0303*} & \underline{0.0473} & \underline{0.0689} & \underline{0.0172} & 0.0226 & \underline{0.0281} \\
&  & IADSR & 0.0297 & \textbf{0.0483*} & \textbf{0.0697*} & \underline{0.0172} & \textbf{0.0230*} & \textbf{0.0287*} \\
\cmidrule{2-9}
& \multicolumn{2}{c|}{STEAM} & 0.0154 & 0.0330 & 0.0630 & 0.0087 & 0.0150 & 0.0214 \\
& \multicolumn{2}{c|}{DCRec} & 0.0204 & 0.0379 & 0.0655 & 0.0123 & 0.0178 & 0.0247 \\
\bottomrule
\end{tabular}
}
\vspace{-3mm}  
\end{table*}

\subsubsection{Evaluation Metrics}
For evaluation, we adopt two highly widely used metrics in sequential recommendation: Hit Ratio (HR@K) and Normalized Discounted Cumulative Gain (NDCG@K). HR@K accurately measures the proportion of test cases where the ground truth item appears in the top-K recommendation list, effectively reflecting the model's ability to recall relevant items~\cite{yang2012top,li2020sampling}. NDCG@K further considers the position of the ground truth item within the top-K list, assigning higher weights to higher positions, thus evaluating both precision and ranking quality~\cite{lim2015top,wang2013theoretical}.
We report results for HR@$K \in\{5, 10, 20\}$, and NDCG@$K \in\{5, 10, 20\}$ to comprehensively evaluate recommendation performance at different levels of K. Following standard practice, we employ the leave-one-out strategy~\cite{black2021leave} for evaluation. 

\subsubsection{Backbones}
Since our method is compatible with different sequential recommendation models, we choose the following three representatives as the backbone~\cite{wang2023plate}.

\begin{itemize}[leftmargin=*]
    \item \textbf{GRU4Rec}~\cite{hidasi2015session}: One of the pioneering works in sequential recommendation that leverages Gated Recurrent Units to effectively capture temporal dynamics in user-item interaction sequences.
        
    \item \textbf{SASRec}~\cite{kang2018self}: This approach introduces self-attention mechanisms into sequential recommendation, enabling the model to adaptively focus on relevant historical interactions while maintaining computational efficiency.
        
    \item \textbf{Caser}~\cite{tang2018personalized}: By employing both horizontal and vertical convolutional filters, this CNN-based method captures local and global sequential patterns simultaneously to enhance accuracy.
\end{itemize}

\subsubsection{Baselines}

We compare our approach with representative denoising sequential recommendation methods, including directly removing noise items and employing data augmentation:

\begin{itemize}[leftmargin=*]
    \item \textbf{STEAM}~\cite{lin2023self}: Self-correcting approach that modifies sequences through keep, delete, or insert operations using self-supervised learning to identify and fix misclicked items.
    
    \item \textbf{DCRec}~\cite{yang2023debiased}: Denoising contrastive framework that separates user conformity from genuine interests using a multi-channel weighting network and contrastive learning.
    
    \item \textbf{HSD}~\cite{zhang2022hierarchical}: Hierarchical sequence denoising model that learns two-level item inconsistency signals to identify and remove noisy interactions without requiring explicit noise labels.

    \item \textbf{SSDRec}~\cite{zhang2024ssdrec}: Framework that uses multi-relation graphs for cross-sequence patterns, injects global information at specific positions, and applies hierarchical denoising to identify noise in both enhanced and original sequences.
\end{itemize}

\subsubsection{Implementation Details}
Following common practices in sequential recommendation, we set the embedding dimension to 64 and the hidden state dimension to 128 with 2 GRU layers. The model is trained with a batch size of 32 and the Adam optimizer with a learning rate of 1e-4. We use the Llama-3.1-8B-Instruct~\cite{grattafiori2024llama} model to generate semantic embeddings with llm2vec~\cite{behnamghader2024llm2vec} and implement early stopping with a patience of 10 epochs to prevent overfitting.

\begin{table*}[ht]
\centering
\tiny  
\setlength\abovecaptionskip{0\baselineskip}
\setlength\belowcaptionskip{0\baselineskip}
\caption{Overall performance comparison on Movielens-100k.}
\label{tab:overall}
\resizebox{0.77\textwidth}{!}{
\begin{tabular}{c|cc|cccccc}
\toprule
Dataset & \multicolumn{2}{c|}{Model} & HR@5 & HR@10 & HR@20 & NDCG@5 & NDCG@10 & NDCG@20 \\
\midrule

\multirow{14}{*}{ML-100K} 
& \multirow{4}{*}{GRU4Rec} & base & 0.0180 & 0.0296 & 0.0607 & 0.0102 & 0.0152 & 0.0230 \\
&  & HSD & 0.0148 & 0.0339 & \textbf{0.0732*} & 0.0094 & 0.0163 & 0.0252 \\
&  & SSDRec & \underline{0.0256} & \textbf{0.0511*} & \underline{0.0721} & \underline{0.0155} & \underline{0.0217} & \underline{0.0268} \\
&  & IADSR & \textbf{0.0286*} & \underline{0.0455} & 0.0696 & \textbf{0.0176*} & \textbf{0.0223*} & \textbf{0.0286*} \\
\cmidrule{2-9}
& \multirow{4}{*}{Caser} & base & 0.0204 & 0.0361 & 0.0541 & 0.0104 & 0.0176 & 0.0210 \\
&  & HSD & \underline{0.0255} & \underline{0.0456} & 0.0721 & \textbf{0.0147*} & \textbf{0.0214*} & 0.0271 \\
&  & SSDRec & 0.0243 & 0.0424 & \underline{0.0732} & \underline{0.0142} & 0.0203 & \textbf{0.0278*} \\
&  & IADSR & \textbf{0.0265*} & \textbf{0.0467*} & \textbf{0.0742*} & \underline{0.0142} & \underline{0.0207} & \underline{0.0276} \\
\cmidrule{2-9}
& \multirow{4}{*}{SASRec} & base & 0.0191 & 0.0350 & 0.0509 & 0.0114 & 0.0153 & 0.0200 \\
&  & HSD & \textbf{0.0223*} & 0.0403 & 0.0742 & \underline{0.0143} & 0.0190 & 0.0256 \\
&  & SSDRec & \textbf{0.0223*} & \underline{0.0435} & \underline{0.0785} & 0.0140 & \textbf{0.0209*} & \underline{0.0295} \\
&  & IADSR & \underline{0.0212} & \textbf{0.0477*} & \textbf{0.0827*} & \textbf{0.0145*} & \underline{0.0205} & \textbf{0.0311*} \\
\cmidrule{2-9}
& \multicolumn{2}{c|}{STEAM} & 0.0207 & 0.0372 & 0.0563 & 0.0126 & 0.0178 & 0.0202 \\
& \multicolumn{2}{c|}{DCRec} & 0.0215 & 0.0424 & 0.0710 & 0.0135 & 0.0202 & 0.0279 \\
\bottomrule
\end{tabular}
}
\vspace{-2mm}  
\end{table*}
\subsection{Overall Performance (RQ1 and RQ2)}

Tables 2 and 3 present the overall performance comparison of our proposed method against baseline and recent denoising models across four datasets. We evaluate IADSR on three backbones (GRU4Rec, Caser, SASRec) and compare it with HSD, SSDRec, as well as standalone frameworks DCRec and STEAM.

From the experimental results, we observe the following findings:

\begin{itemize}[leftmargin=*]
    \item \textbf{Effectiveness of Our Approach:} IADSR consistently outperforms the second-best approaches across different backbones and datasets. On Beauty with GRU4Rec, compared to the second-best method (HSD), IADSR shows average improvements of 24.6\% across all metrics (31.0\% on HR@5, 8.5\% on HR@10, while HSD leads on HR@20, 51.1\% on NDCG@5, 31.9\% on NDCG@10, and 14.1\% on NDCG@20). On Toys with Caser, the average gain reaches 36.3\%. On ML-100K with SASRec, IADSR achieves a 7.8\% average gain across metrics. These consistent gains highlight its robust denoising capability.
    
    \item \textbf{Performance Across Different Backbones:} Improvements hold regardless of the backbone. Even with SASRec, the strongest baseline, IADSR raises HR@20 on Beauty from 0.0554 to 0.0836 (+50.9
    
    \item \textbf{Comparison with Other Denoising Methods:} IADSR outperforms HSD and SSDRec on most metrics. While SSDRec occasionally excels (e.g., HR@20 with GRU4Rec on Beauty), IADSR provides more balanced improvements, achieving average HR@10 gains of 10.2\%, 8.4\%, and 6.7\% on Beauty, Sports, and Toys, respectively.
    
    \item \textbf{Dataset-specific Observations:} Gains are most pronounced on Beauty and Sports, suggesting higher noise levels. On ML-100K, improvements are smaller, implying less or different noise patterns.
\end{itemize}

Our denoising framework consistently outperforms state-of-the-art methods across multiple backbones by integrating semantic and collaborative signals to effectively mitigate noise in sequential recommendation, demonstrating robust and versatile performance across datasets.

\subsection{Ablation Study (RQ3)}

To validate the effectiveness of each component in our proposed framework, we conduct an ablation study on the Beauty dataset. Table 4 presents the results with different variants of our model. Specifically, we investigate the impact of different loss functions and interest embeddings:

\begin{itemize}[leftmargin=*]
   \item \textbf{w/o ${both}$}: Removing both InfoNCE loss and reconstruction loss, leaving only the basic cross-entropy loss.
   \item \textbf{w/o $\mathcal{L}_{info}$}: Removing the InfoNCE loss that aligns with semantic and collaborative embeddings.
   \item \textbf{w/o $\mathcal{L}_{recon}$}: Removing the sequence reconstruction loss.
   \item \textbf{Short-only}: Using only short-term interests for noise detection.
   \item \textbf{Long-only}: Using only long-term interests for noise detection.
   \item \textbf{Full Model}: Our full model with all components.
\end{itemize}

The results demonstrate that removing both losses leads to substantial drops (–42.0\% HR@5, –38.8\% NDCG@5), confirming their importance. $\mathcal{L}{info}$ proves more critical than $\mathcal{L}{recon}$, highlighting the necessity of aligning semantic and collaborative spaces.

Regarding interest embeddings, we observe that long-term interests offer more stable signals than short-term ones, but the Full Model outperforms both (with improvements of up to 33.3\% in HR@5 and 36.1\% in NDCG@5 compared to Long-only), confirming our hypothesis that combining both time scales offers the most comprehensive view for identifying noise in user sequences.

\subsection{Hyper-parameter Study (RQ4)}

To understand the impact of the cross-modal consistency threshold $\theta$ on our framework's performance, we conducted experiments varying this parameter from -1.0 to 0.9. Figure 3 illustrates performance trends across different metrics. We observe that performance peaks around $\theta = -0.9$ (with HR@5=0.03 and NDCG@5=0.0213) and remains relatively stable across negative thresholds (-0.9 to -0.1). However, as $\theta$ exceeds 0.7, we observe significant performance degradation across all metrics.

These results indicate that moderate cross-modal alignment is sufficient for effective noise identification. Setting $\theta$ too high forces excessive agreement between semantic and collaborative signals, potentially ignoring complementary information. Based on these findings, we set $\theta = -0.9$ as the default value in our framework.

\begin{table}[t]
\small
\setlength\abovecaptionskip{0\baselineskip}
\setlength\belowcaptionskip{0\baselineskip}
\centering
\setlength{\tabcolsep}{2pt} 
\caption{Ablation study on the Amazon Beauty dataset.}
\label{tab:ablation}
\begin{minipage}{0.48\textwidth} 
\begin{tabular}{lcccccc}
\toprule
\textbf{Variants} & \textbf{HR@5} & \textbf{HR@10} & \textbf{HR@20} & \textbf{NDCG5} & \textbf{NDCG10} & \textbf{NDCG20} \\
\midrule
w/o ${both}$ & 0.0174 & 0.0249 & 0.0351 & 0.0120 & 0.0144 & 0.0169 \\
w/o $\mathcal{L}_{info}$ & 0.0283 & 0.0346 & 0.0386 & 0.0202 & 0.0223 & 0.0233 \\
w/o $\mathcal{L}_{recon}$ & 0.0213 & 0.0305 & 0.0416 & 0.0139 & 0.0168 & 0.0196 \\

\midrule
Short-only & 0.0218 & 0.0329 & 0.0483 & 0.0144 & 0.0180 & 0.0219 \\
Long-only & 0.0225 & 0.0328 & 0.0456 & 0.0151 & 0.0181 & 0.0223\\
\textbf{Full Model} & \textbf{0.0300} & \textbf{0.0396} & \textbf{0.0486} & \textbf{0.0196} & \textbf{0.0259} & \textbf{0.0321} \\
\bottomrule
\end{tabular}
\end{minipage}
\end{table}

\begin{figure}[t]
\small
\setlength\abovecaptionskip{0.5\baselineskip}
\setlength\belowcaptionskip{0.5\baselineskip}
	\centering
	\begin{minipage}{0.495\linewidth}
		\centering
        \begin{subfigure}{1\linewidth}
		\includegraphics[width=0.995\linewidth]{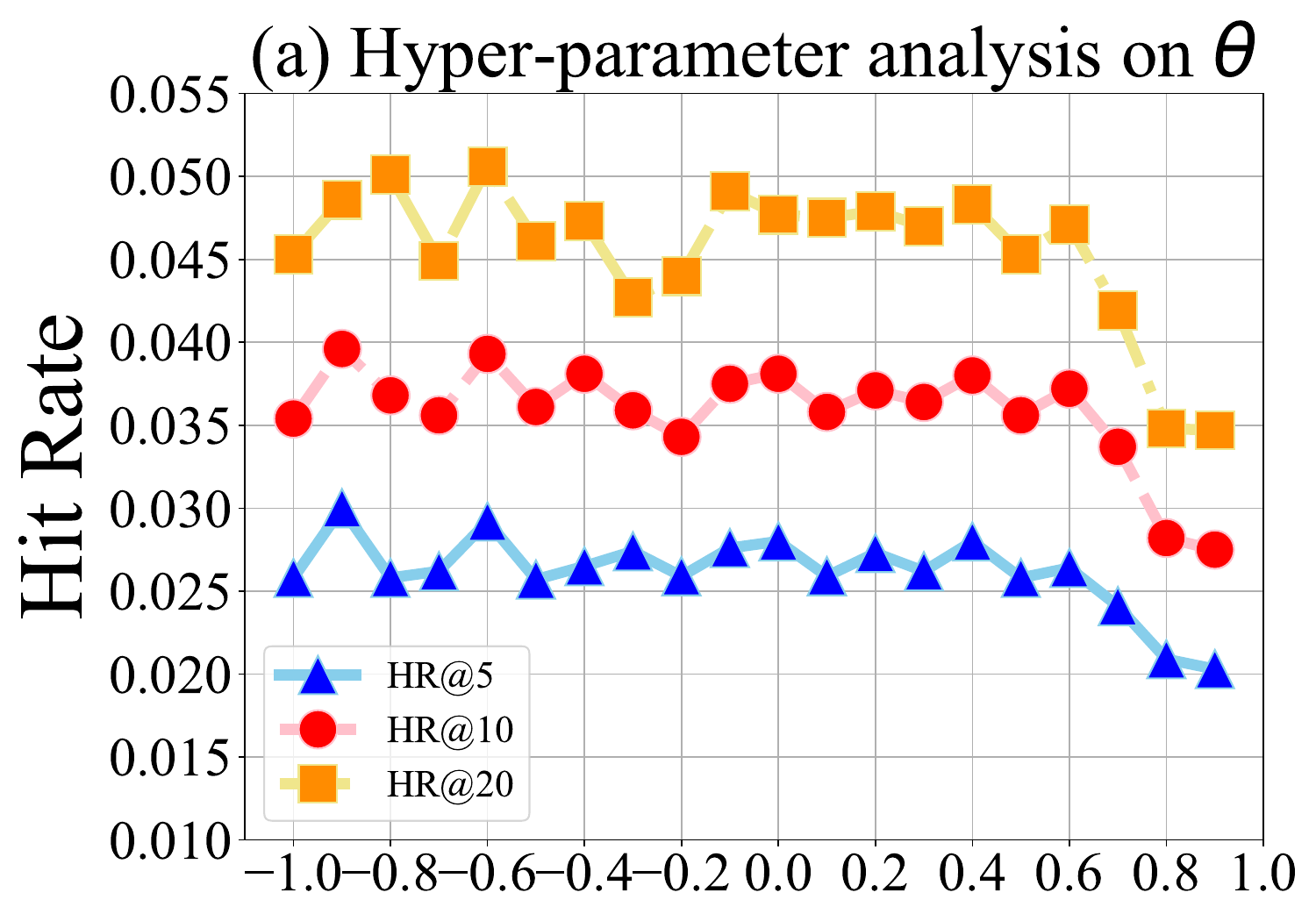}
		\label{fig:RQ2-1}
        \end{subfigure}
	\end{minipage}
	\begin{minipage}{0.495\linewidth}
		\centering
        \begin{subfigure}{1\linewidth}
		\includegraphics[width=0.995\linewidth]{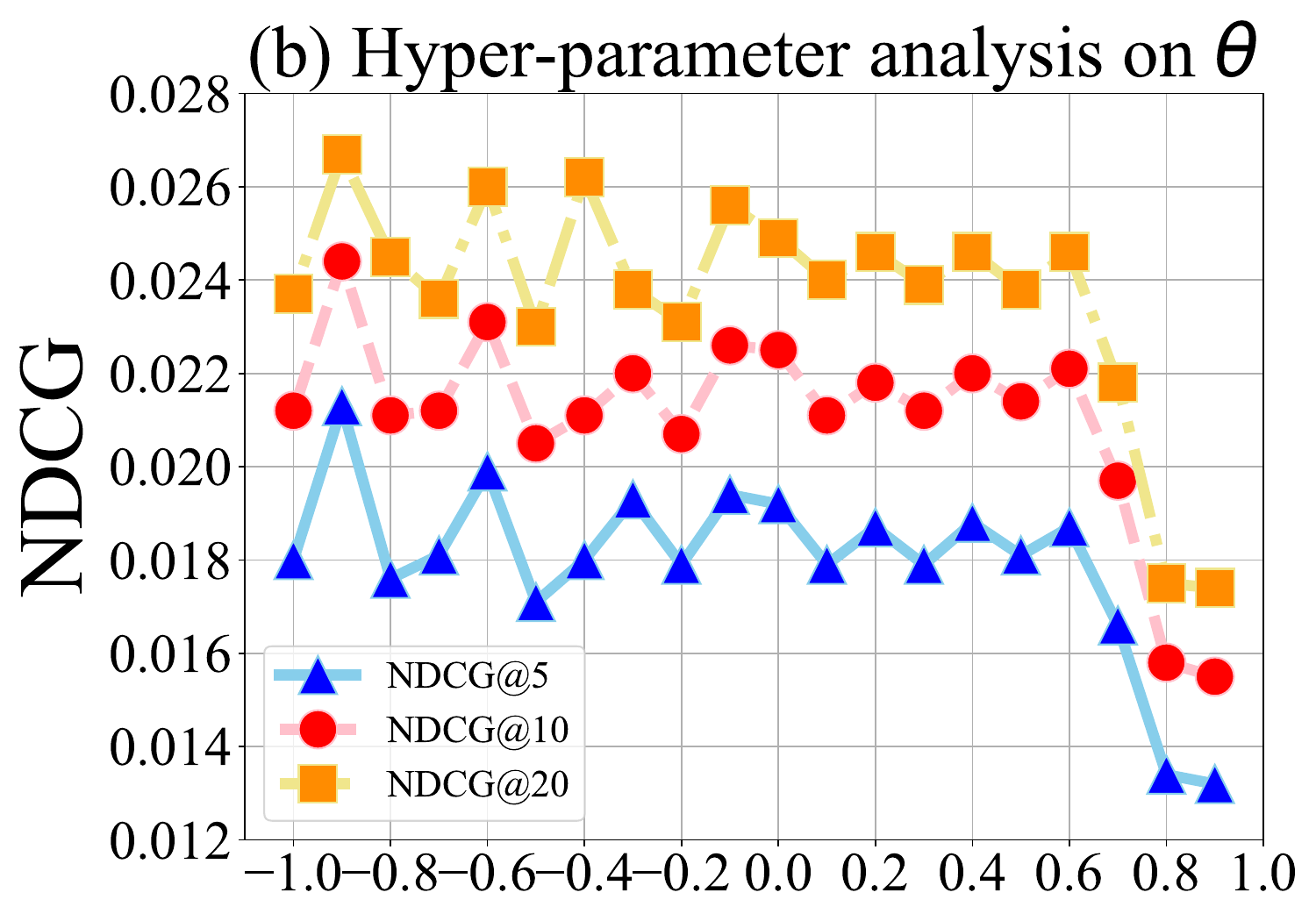}
		\label{fig:RQ2-2}
        \end{subfigure}
	\end{minipage}
	\vspace{-5mm}
	\caption{(a) HR@5 and (b) NDCG@5 in  hyper-parameter study on $\theta$.} 
        \vspace{-8mm}
	\label{fig:RQ2}
\end{figure}

\subsection{Case Study (RQ5)}

To provide qualitative insights into our model's denoising capability, we randomly selected two users from the Beauty dataset and analyzed how our approach effectively identifies noisy interactions by leveraging both semantic and collaborative signals. 

Table 5 illustrates the denoising results across different methods (IADSR, HSD, Steam) for these selected users. The blue text indicates cold items (representing the 20\% of items with the lowest interaction counts), the red text represents hot items (the 20\% with the highest interaction counts), and the black text denotes normal items, respectively. Furthermore, we display the user profile derived from their historical interaction patterns.

\begin{table*}[htbp]
\centering
\small
\caption{A case study to demonstrate the effectiveness of IADSR in denoising cold items.}
\vspace{-6pt}
\renewcommand{\arraystretch}{0.9}
\setlength{\tabcolsep}{0.8pt}
\begin{tabular}{|p{1.3cm}|p{4cm}|p{4cm}|p{4cm}|p{4cm}|}
\hline
\textbf{User ID} & \textbf{IADSR} & \textbf{HSD} & \textbf{Steam} & \textbf{User Profile} \\
\hline
\vspace{1pt}22099 & 
\begin{tabular}[t]{@{}p{1.8cm}p{2.0cm}@{}}
\parbox[t][1cm][t]{1.4cm}{\vspace{1pt}\textcolor{cyan}{Bedhead Hook Up 1/2 STRAIGHT}} & 
\parbox[t][1cm][t]{1.6cm}{\vspace{1pt}\centering\includegraphics[width=0.7cm]{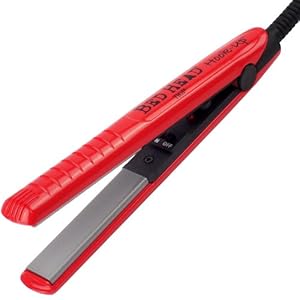}}
\end{tabular}
& 
\begin{tabular}[t]{@{}p{1.8cm}p{2.0cm}@{}}
\parbox[t][1cm][t]{1.4cm}{\vspace{1pt}\textcolor{cyan}{Bedhead Hook Up 1/2 STRAIGHT}} & 
\parbox[t][1cm][t]{1.6cm}{\vspace{1pt}\centering\includegraphics[width=0.7cm]{case_study/Bedhead.jpg}} \\[0.8cm] 
\parbox[t][1cm][t]{1.8cm}{\vspace{1pt}\textcolor{red}{Skinny Cream Clinically Proven Cellulite Reduction, 6 Ounce}} &
\parbox[t][1cm][t]{2.0cm}{\vspace{1pt}\centering\includegraphics[width=0.6cm]{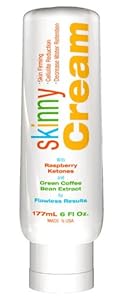}}
\end{tabular}
& 
\begin{tabular}[t]{@{}p{1.8cm}p{2.0cm}@{}}
\parbox[t][0.8cm][t]{1.8cm}{\vspace{1pt}Slim Extreme 3d Super-concentrated Serum Shaping Buttocks, 200mL} & 
\parbox[t][0.8cm][t]{2.0cm}{\vspace{1pt}\centering\includegraphics[width=0.6cm]{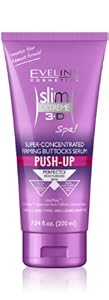}} \\[2cm]
\parbox[t][0.8cm][t]{1.8cm}{\vspace{1pt}\textcolor{cyan}{Palmers Cocoa Butter Bust Firming Cream 4.4oz}} &
\parbox[t][0.8cm][t]{2.0cm}{\vspace{1pt}\centering\includegraphics[width=1cm]{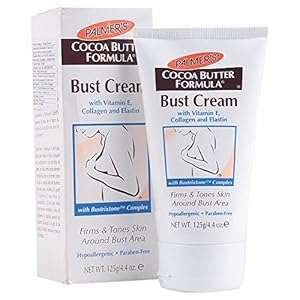}}
\end{tabular} & 
\parbox[t][4.5cm][t]{4cm}{\vspace{1pt}Based on the user's historical interactions, this user is primarily interested in body contouring, firming, and weight loss-related beauty products, with a particular focus on shaping creams and anti-cellulite products targeting specific body areas (buttocks, bust).}\\[-0.6cm]
\hline
\vspace{1pt}19852 & 
\begin{tabular}[t]{@{}p{1.8cm}p{2.0cm}@{}}
\parbox[t][1.8cm][t]{1.8cm}{\vspace{1pt}\textcolor{cyan}{Beauty Without Cruelty Fragrance Free Hand \& Body Lotion, 100 \% Vegetarian, 16 fl ozs}} & 
\parbox[t][1.8cm][t]{2.0cm}{\vspace{1pt}\centering\includegraphics[width=0.7cm]{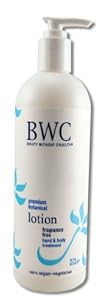}}
\end{tabular} & 
\begin{tabular}[t]{@{}p{1.8cm}p{2.0cm}@{}}
\parbox[t][2.5cm][t]{1.8cm}{\vspace{1pt}\textcolor{cyan}{Beauty Without Cruelty Fragrance Free Hand \& Body Lotion, 100 \% Vegetarian, 16 fl ozs}} & 
\parbox[t][2.5cm][t]{2.0cm}{\vspace{1pt}\centering\includegraphics[width=0.7cm]{case_study/BeautyFree.jpg}} \\[2.4cm] 
\parbox[t][1.8cm][t]{1.8cm}{\vspace{1pt}\textcolor{cyan}{Essie Ridge Filler Base Coat, 0.46 oz}} & 
\parbox[t][1.8cm][t]{2.0cm}{\vspace{1pt}\centering\includegraphics[width=1cm]{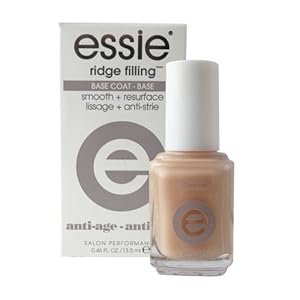}} \\[-0.8cm] 
\parbox[t][1.8cm][t]{2.0cm}{\vspace{1pt}\textcolor{cyan}{Fruit Of The Earth 100\% Aloe Vera 6oz. Gel Tube}} & 
\parbox[t][1.8cm][t]{2.0cm}{\vspace{1pt}\centering\includegraphics[width=1cm]{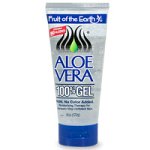}}
\end{tabular} &
\begin{tabular}[t]{@{}p{1.8cm}p{2.0cm}@{}}
\parbox[t][2.5cm][t]{1.8cm}{\vspace{1pt}\textcolor{red}{Hydroxatone AM/PM Anti-Wrinkle Complex SPF 15}} & 
\parbox[t][2.5cm][t]{2.0cm}{\vspace{1pt}\centering\includegraphics[width=1.2cm]{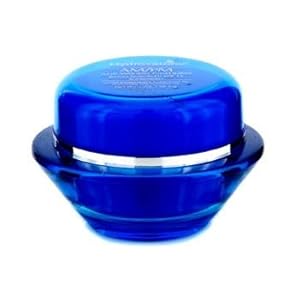}} \\[-0.8cm] 
\parbox[t][1.8cm][t]{1.8cm}{\vspace{1pt}\textcolor{cyan}{Blinc Kiss Me Mascara, Dark Brown}} & 
\parbox[t][1.8cm][t]{2.0cm}{\vspace{1pt}\centering\includegraphics[width=1.2cm]{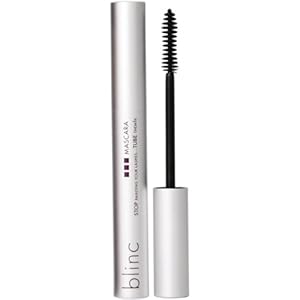}} \\[-0.5cm] 
\parbox[t][1.8cm][t]{1.8cm}{\vspace{1pt}\textcolor{cyan}{Alkaline}} & 
\parbox[t][1.8cm][t]{2.0cm}{\vspace{1pt}\centering\includegraphics[width=1.2cm]{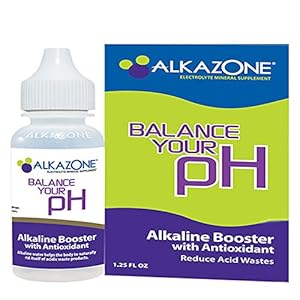}}
\end{tabular} & 
\parbox[t][3cm][t]{3.9cm}{\vspace{1pt}Based on the user's historical interactions, this user has comprehensive beauty interests across multiple categories, showing particular focus on anti-aging skincare, natural/vegetarian beauty products, eye makeup, hair care tools, and various face and body treatments from both high-end and drugstore brands.} \\[-0cm]
\hline
\end{tabular}
\vspace{-4mm}
\label{tab:casestyudy}
\end{table*}

\subsubsection{User Preference Analysis}
For User 22099, IADSR precisely filtered the irrelevant hair styling product while retaining body-contouring items, outperforming HSD and STEAM, which misclassified relevant products.

For User 19852, IADSR flagged only the “Beauty Without Cruelty” lotion as noise, while preserving other cold items consistent with the user’s beauty interests. HSD overgeneralized by filtering all three natural products, including relevant ones such as the Essie base coat and Aloe Vera gel. STEAM performed worse, incorrectly marking an anti-wrinkle complex and eye makeup as noise despite the user’s clear interest in anti-aging skincare and eye products. These cases demonstrate IADSR’s ability to balance recommendation diversity with precise noise filtering, ensuring both relevance and coverage.



\section{Related Work}  \label{sec:relatedwork}
In this section, we summarize the related works on sequential recommender systems and denoising sequential recommendation.

\subsection{Sequential Recommender System}

The sequential recommendation focuses on capturing temporal dynamics in user behaviors to predict future interactions. Early approaches used Markov Chain-based models~\cite{rendle2010factorizing}, which were later superseded by deep learning methods~\cite{lin2023autodenoise}, including GRU4Rec~\cite{hidasi2015session} with RNNs and Caser~\cite{tang2018personalized} with CNNs. Recently, Transformer-based models like SASRec~\cite{kang2018self} have achieved consistently superior performance by leveraging self-attention mechanisms to model item relationships in user sequences.

However, these models lack noise mitigation mechanisms, making them vulnerable to accidental clicks or exploratory behaviors. IADSR addresses this by integrating LLM semantic knowledge with collaborative signals to distinguish preferences from noise.

\subsection{Denoising Sequential Recommendation}

Recent works on denoising sequential recommendation can be grouped into two categories.
The first type relies solely on collaborative signals (ID-based interactions). For example, HSD~\cite{zhang2022hierarchical} detects inconsistency signals to drop noisy items, while ADT~\cite{wang2021denoising} prunes high-loss interactions during training. Other approaches, such as STEAM~\cite{lin2023self}, SSDRec~\cite{zhang2024ssdrec}, DCRec~\cite{yang2023debiased}, and DCF~\cite{he2024double}, adjust or reweight noisy items through self-correction, graph modeling, or debiased contrastive learning.
The second type incorporates additional modalities beyond IDs. LLM4DSR~\cite{wang2024llm4dsr} leverages large language models to identify and replace noisy items, while LLaRD~\cite{wang2025unleashing} extracts semantic preference patterns from textual contexts.

While ID-based methods risk overlooking the semantic richness of user behaviors, multimodal approaches struggle to align heterogeneous signals with collaborative sequences. Existing denoising strategies also face notable limitations: (1) reliance on collaborative signals makes them ineffective for cold items with sparse interactions; (2) LLM-based methods often demand costly fine-tuning; and (3) many are tied to specific architectures, limiting generalizability. In contrast, our framework leverages LLM embeddings to improve denoising, remains compatible with diverse sequential recommenders, and is particularly effective for cold items where collaborative signals are insufficient.

\section{Conclusion} \label{sec:conclusion}

In this paper, we proposed IADSR, a novel framework that integrates semantic knowledge from large language models with collaborative signals for denoising sequential recommendation. Through a two-stage process of cross-modal alignment, noise detection, and sequence reconstruction, IADSR effectively preserves real user preferences. Experiments on four public datasets show that it consistently outperforms state-of-the-art denoising methods across different sequential recommendation backbones.

\section*{ACKNOWLEDGEMENT}

This research was partially supported by Hong Kong Research Grants Council's Research Impact Fund (No.R1015-23), Collaborative Research Fund (No.C1043-24GF), General Research Fund (No.11218325), Institute of Digital Medicine of City University of Hong Kong (No.9229503), Huawei (Huawei Innovation Research Program), Tencent (CCF-Tencent Open Fund, Tencent Rhino-Bird Focused Research Program), Alibaba (CCF-Alimama Tech Kangaroo Fund No. 2024002), Ant Group (CCF-Ant Research Fund), Didi (CCF-Didi Gaia Scholars Research Fund), Kuaishou, Bytedance, and National Natural Science Foundation of China (No.62502404).

\section*{GenAI Usage Disclosure}

We provide full disclosure of our use of GenAI tools throughout this research and writing:

\textbf{Data Processing}: We utilized Claude to 
transform raw datasets into the required format during the data processing stage, ensuring data consistency and reducing manual processing errors.


\textbf{Writing}: Claude was employed solely for grammar checking and improving sentence clarity and expression. 


\bibliographystyle{ACM-Reference-Format}
\balance
\bibliography{7-reference}

\end{document}